\documentclass[submission,copyright,creativecommons]{eptcs}
\providecommand{\event}{WFLP 2016} % Name of the event you are submitting to
\usepackage{breakurl}             % Not needed if you use pdflatex only.

\usepackage[english]{babel}
\usepackage{amsmath}
\usepackage{mathtools}[2011/02/12]
\usepackage{amsfonts,amssymb}
\let\vec\relax
\usepackage{MnSymbol}
\usepackage{mdwlist}
\usepackage{multicol}
\usepackage{galois}
\usepackage{graphicx}
\usepackage{caption}
\usepackage[flushleft]{threeparttable}
\usepackage{tikz}
\usetikzlibrary{arrows,automata}

\usepackage[utf8]{inputenc}

\include{wgmacros-TCCP}
\usepackage{listings}		% Source code listings
\definecolor{dkgreen}{rgb}{0,0.6,0}
\definecolor{gray}{rgb}{0.5,0.5,0.5}
\definecolor{mauve}{rgb}{0.58,0,0.82}
\definecolor{orange}{rgb}{1,0.5,0}

\newcommand{\ppl}{{\sc PPL}}

\newcommand*{\java}{Java}

\DeclareMathSizes{10}{10}{8}{8}   % For size 10 text

\newenvironment{Draftpleasenote*}[1][??]{\begin{draftpleasenote*}[{#1}]}{\end{draftpleasenote*}}

\usepackage[fancyproofs,noextended,squareitemtag]{theorems}[2013/01/24]%,fancyexamples,final

\newcommand*{\hytccp}{{\textsc hy}-\-\tccp}

\renewcommand*{\askip}{\textsf{stop}}
\renewcommand*{\CSmerge}{\wedge}

\newcommand*{\Confc}{\mathit{Conf}}

\monobioperator{\Ldisj}{\mathrel{\dot\vee}}{\mathrel{\dot\bigvee}}
\monobioperator{\Lconj}{\mathrel{\dot\wedge}}{\mathrel{\dot\bigwedge}}

\newcommand{\lista}[3]{#1=[#2$\mid$#3]}

\newcommand{\ask}{\textsf{ask}}
\newcommand{\tell}{\textsf{tell}}

\usepackage{amsmath}

\title{A Simulation Tool for \tccp\ Programs\thanks{This work has been supported by the Andalusian Excellence Project P11-TIC7659.}}

\author{Mar\'ia-del-Mar Gallardo \and Leticia Lavado \and Laura Panizo
           \institute{Universidad de M\'alaga, Andaluc\'ia Tech, Dept. Lenguajes y Ciencias de la Computaci\'on, Espa\~{n}a.}
           \email{[gallardo,leticialavmu,laurapanizo]@lcc.uma.es }}

\def\titlerunning{A Simulation Tool for \tccp}
\def\authorrunning{M.M.Gallardo, L. Lavado, \& L. Panizo}
\begin{document}
\maketitle

\begin{abstract}
The Timed Concurrent Constraint Language \tccp\ is a declarative synchronous concurrent language, particularly suitable for modelling reactive systems.
In \tccp, agents communicate and synchronise through a global constraint store.
%that is monotonic in the sense that added constraints cannot never be removed.
It supports a notion of discrete time that allows all non-blocked agents to proceed with their execution simultaneously.

%In this paper, we present an abstract machine-based tool for the execution of \tccp\ programs. The abstract machine has the following  components.
%First, the instructions and memory model to represent the agents and constraints. Second, the execution model of the programs . Finally, the constraint solver components which are the external modules that deal with the constraints.

%In this paper, we present an abstract machine-based architecture for the simulation of \tccp\ programs. The abstract machine has three main components.
In this paper, we present a modular architecture for the simulation of \tccp\ programs. The tool comprises three main components.
First, a set of basic abstract instructions able to model the \tccp\ agent behaviour, the memory model needed to manage the active agents and the state of the store during the execution. Second,  the agent interpreter that executes the instructions of the current agent iteratively and calculates the new agents to be executed at the next time instant. Finally, the constraint solver components which are the  modules that deal with constraints.
%First, the instructions and memory model  needed to  represent the active agents and the state of the store during the execution. Second,  the agent interpreter that executes the instructions of the current agent iteratively and calculates the new agents to be executed at the next time instant. Finally, the constraint solver components which are the  modules that deal with the constraints. %Specifically, we have used an ad-hoc solver to deal with the logic constraints, and the Parma Polyhedra Library to handle the linear constraints.

In this paper, we describe the implementation of these components and present an example of a real system modelled in \tccp.

\end{abstract}

\section{Introduction}
\label{sec:intro}
It is well known that many critical applications in different domains,
such as health~\cite{Tuan10}, railways~\cite{Qian15} or automotive~\cite{Kang13} have a reactive and concurrent behaviour that is
difficult to model and analyse. Unfortunately, certain errors in these applications may have highly negative consequences and, therefore, it is essential to detect failures in software in the early design phases. This is why most modelling languages for these complex systems are supported by simulation and verification tools that guarantee the software's safety and reliability with respect to the critical properties.

%Nowadays, reactive and concurrent systems allow us to model many critical applications in different domains,
%such as health~\cite{Tuan10}, railways~\cite{Qian15} or automotive~\cite{Kang13}.
%% Laura - comentado
%Since the tasks carried out by these systems may be critical, an error
%%in this kind of systems
%can lead to negative consequences.
%%Laura - añadido

%Different  languages have been proposed for the modelling of these systems

%In critical applications, an error can have negative consequences.
%For this reason, it is essential to guarantee the system safety and reliability.
%Unfortunately, the description, verification and analysis of concurrent software are very hard tasks owing to the concurrent execution and synchronisation.

% Leti - hybrid systems
%Hybrid systems are a particular type of reactive and concurrent systems that combine continuous and discrete components.
%Continuous behaviours usually describe the evolution of physical variables, whereas discrete behaviours model the different execution modes of the system. The combination of these two aspects complicates the design, validation and analysing of these applications.

Several formalisms have been developed to solve the problem of describing and analysing concurrent systems.
In this paper, we focus on the Concurrent Constraint Paradigm (\ccp)~\cite{SaraswatR90} characterised by the use of store-as-constraint instead of the classical store-as-value paradigm. Specifically, \tccp~\cite{deBoerGM99} is a language suitable for describing reactive systems within this paradigm. As opposed to the interleaving composition of processes supported by most concurrent modelling languages, \tccp\ makes use of the synchronous composition of processes. Synchronous languages have proved to be very useful for modelling hardware and software systems. Some successful examples are {\em Lustre}~\cite{Ha91} or {\em SIGNAL}~\cite{GautierG87}. The synchronous management of processes clearly simplifies the scheduling tasks, although it complicates the memory use. The declarative and synchronous character of \tccp\ makes it particularly suitable not only for describing but also for analysing complex concurrent systems.

There are a few tools for \tccp\ proposed in the literature~\cite{Lescaylle200963, Sjoland01}. In this paper,
we present a modular framework for \tccp\ with the aim of overcoming the lack of simulation and analysis tools.  
%The proposal has a modular architecture  based on the definition of an abstract machine-based  for the construction of tools for the simulation and analysis of \tccp.
Classically, the implementation of logic languages has been based on the definition of the so-called {\em abstract machines} which provide an abstraction layer on the ultimate device that
will execute the programs. Warren Abstract Machine~(WAM)~\cite{WAM91} is the first and most well-known abstract machine for logic languages. In the context of concurrent logic languages, there exist other proposals such as the abstract machine based on the construction of an \textsf{AND}/\textsf{OR} tree for the implementation of \textit{Parlog}~\cite{Gregory1989}, or the Parallel Inference Machine (PMI) for language {\em KL1}~\cite{Ueda1990}.

We have built a simulation tool for \tccp\ programs following the abstract machine philosophy but with some differences derived from the special features of the language\footnote{The prototype tool can be found at \url{http://morse.uma.es/tools/tccp}. }. The construction of a tool for executing \tccp\ implies dealing with its declarative,  constraint-based and synchronous character. %, and hybrid character
For instance, the logic and concurrent nature of \tccp\ involves the creation of a large number of fine-grain agents with a well-delimited variable scope. In addition, the use of constraints as data makes the integration of constraint solvers in the tool necessary. Furthermore, to correctly deal with the synchronisation, all agents executing in parallel must have a consistent view of the global memory (called {\em store} in \tccp).
%Finally, the hybrid nature of \hytccp\ means that the store has to be divided into two memory zones: the discrete zone that behaves monotonically as in the original \tccp\ language, and the continuous zone that can not be monotonic since continuous variables must change their value. The coexistence of these two zones in the store clearly complicates the language's implementation.
%discrete/continuous stores clearly complicates the language implementation.

The implemented tool comprises different components
%that behave as an abstract machine and are oriented
to successfully solve the aforementioned problems. The core of the tool is formed by a set of abstract instructions and a memory model that is able to represent the state of the \tccp\ program (that is, the current agent and the state of the global store) during the execution. In addition, the tool includes an interpreter that executes the current active agent iteratively. Finally,
%the abstract machine has
there is a module with the constraint solvers used to manage the basic operations on the global store correctly. %It is worth noting that the implementation of the continuous evolution of agents is based on a discretized operational semantics of \tccp\ which is a correct abstraction of the operational semantics that uses convex polyhedra to represent the values reached by the discrete variables. %For this reason, we restrict continuous variables values and their flows to rectangular sets. Due to its extensiveness, it has not been possible to include this discretized semantics in the paper.

%It is clear that support tools are needed to make any formalism useful. The development of tools based on abstract machines is an interesting approach,
%since the abstract machine allows a modular implementation that is independent of the language semantics.
%In this work we define an abstract machine for \hytccp. We have used the as a reference. The WAM, consisting of a memory model and a set of instructions, has as purpose to make interpretation of Prolog program more efficient. Follow that model, the \hytccp\ abstract machine has
%to construct the \hytccp\ abstract machine.
%on the one hand, the memory model which we call abstract store represents the global structure of the memory used when a \hytccp\ program is running.
%On the other hand, the set of abstract instructions manages the abstract store and makes possible the compilation of the language.
%Finally, we are developing a prototype simulator for \hytccp\ based on the abstract machine. Due to \hytccp\ semantics are an extension of \tccp\,
%the abstract machine and the tool accepts also \tccp\ programs.
In this paper, we describe all these components, their implementation and evaluation with a typical \tccp\ example. %It is worth noting that the way that the basic instructions are defined in the abstract machines makes it possible to reason about the correctness of the implementation.

The rest of the paper is organised as follows. Section~\ref{sec:background} presents \tccp\ language syntax and semantics. Section~\ref{sec:approach} describes the approach and the main elements of the abstract machine: the memory model, the instructions and the agent interpreter. In Section~\ref{sec:implementation}, we comment on some implementation issues of the prototype simulator. Section~\ref{sec:evaluation} shows the simulation of the \tccp\ example, with several results and measures obtained after different executions. In Section~\ref{sec:related}, we present some related work. And finally, in Section~\ref{sec:conclusions}, we present the conclusions and future work. 

\section{Introducing \tccp}
\label{sec:background}
%\subsection{Introducing \tccp}

%Several formalisms have been developed to solve the problem of describing and analyzing concurrent systems.
As stated in the Introduction, in this paper, we focus on the Concurrent Constraint Paradigm (\ccp)~\cite{SaraswatR90} which is characterised by the use of store-as-constraint instead of the classical store-as-value paradigm. Within this paradigm, \tccp~\cite{deBoerGM99} is a well-known language suitable for describing reactive systems.
In \tccp, agents execute synchronously in parallel, and communicate across a global monotonic constraint  {\em store}.
% Laura - añadido
The store is monotonic in the sense that the constraints added can never be removed.
%Laura - comentado
%. The monotony of the store means that the constraints added can not be ever removed.
The language includes the notion of discrete time and the capability to capture the absence of information.

%
%Language \hytccp{} was introduced in \cite{AdalidGT14,AdalidGT15} as a conservative extension of
%%the Timed Concurrent Constraint Programming language (
%\tccp~\cite{deBoerGM99}
%to model hybrid systems.
%It inherits from \tccp{} the concurrency and synchronization features,
%which provides an expressive model for the discrete behavior of hybrid systems.
%Furthermore, \hytccp{} includes primitives that are able to handle continuous time
%and the evolution of continuous variables.
%
%In \hytccp{}, continuous time
%is defined
%by means of a global continuous clock. The computation
%proceeds as the concurrent execution of some agents
%that communicate and synchronize
%by means of a \emph{global store}.
%This store is composed of a monotonic \emph{discrete store}
%(once we add information in it, it cannot be deleted), which is a constraint
%%containing informations about
%modeling
%the current state
%and the synchronization signals,
%and a non-\-monotonic \emph{continuous store}
%containing information
%about the continuous variables.

%
The \tccp\ language is parametric \wrt{}
%a cylindric
%constraint system.
% and to a family of ODEs.
%  that are used to model the discrete
% and the continuous information of the system, \resp.
%Briefly,
a \emph{cylindric
constraint system} that is able to abstractly capture the notion of shared constraint store over which two main operations can be carried out.
The {\em write} operation (denoted as \tell\ in the language) that updates the  store with new constraints, and the {\em read} operation (denoted as \ask\ in the language)  to request whether a given constraint is {\em entailed} by the store.
% A constraint system can be seen
% as a partial information system, \ie{} instead of knowing
% the specific value of a
% variable, just partial information
% %on this variable
% is available. In other words, the lack of information about a variable $x$
% does not imply that $x=2$ nor that $x\neq2$ since no information is available at all.

% Thus, as it happens in all the \ccp{} extensions, in \hytccp{} the notion
% of \emph{store-as-valuation} from von Neumann is replaced with
% the notion of \emph{store-as-constraint}.

\begin{definition}[Cylindric constraint system \cite{deBoerGM99}]\label{def:CylCS} \mbox{}
%\index{cylindric constraint system}
    A  cylindric constraint system is an algebraic structure of the
    form $\CSys= \langle \CSdom,\, \CSord,\, \CSmerge,\, \CStrue,\, \CSfalse,\, \Var,\, \exists \rangle $
    such that:
    \begin{enumerate}
        \item $\langle \CSdom,\, \CSord,\, \CSmerge,\, \CStrue,\, \CSfalse \rangle$
        is a complete lattice where
        $\CSmerge$ is the least upper bound ($lub$)
        operator, and
        $\CStrue$ and $\CSfalse$ are, \resp, the least and the greatest
        elements of $\CSdom$. We often
        use the inverse order $\vdash$ (the \emph{entailment} relation) instead
        of $\CSord$ over constraints.  Formally $\forall c,d\ \in \CSdom$
        $c \CSord d\Leftrightarrow d \CSimp c$.
        \item $\Var$ is a
        denumerable set of variables.
        \item For each element $x\in \Var$, a
        function (called cylindric operator) $\exists_{x} \colon
        \CSdom\ra \CSdom$ is defined such that, for any $c,d\in \CSdom$ the
        following axioms hold:% (i) $c\CSimp \exists_{x}c$,
        %(ii) if $c\CSimp d$ then $\exists_{x}c\CSimp \exists_{x}d$,
        %(iii) $\exists_{x}(c\CSmerge \exists_{x}d)=\exists_{x}c\CSmerge \exists_{x}d$,
       % and
       % (iv) $\exists_{x}(\exists_{y}c)=\exists_{y}(\exists_{x}c)$.
         \begin{enumerate}
           \item $c\CSimp \exists_{x}c$
          \item if $c\CSimp d$ then $\exists_{x}c\CSimp
           \exists_{x}d$
          \item $\exists_{x}(c\CSmerge
          \exists_{x}d)=\exists_{x}c\CSmerge
           \exists_{x}d$
           \item
          $\exists_{x}(\exists_{y}c)=\exists_{y}(\exists_{x}c)$
           \item To model parameter passing, \emph{diagonal
        elements} are added to the primitive
        constraints. For all $x$, $y$ ranging over $\Var$,
        the constraint $d_{xy}$ which satisfies the
         following axioms is added.
         \begin{enumerate}
             \item $\CStrue\CSimp d_{xx}$
             \item if $z \neq x,y$ then
             $d_{xy}=\exists_{z}(d_{xz}\CSmerge d_{zy})$
             \item if $x\neq y$ then
             $\exists_{xy}(c\CSmerge d_{xy})\CSimp c$.
         \end{enumerate}
         Diagonal elements represent the equality relation between variables in the constraint systems.
        \end{enumerate}
    \end{enumerate}
\end{definition}

The syntax of \tccp{} agents is given by the following grammar:
%\vspace{-1.7ex}
\begin{align*}
    A ::=\, &\askip \mid \atell{c} \mid A \parallel A \mid
    \anow{c}{A}{A} \mid \ahiding{x}{A} \mid p(\vec{x}) \mid  \textstyle{\textstyle{\sum_{i=1}^{n}\aask{c_{i}} A }}
\end{align*}
%
%\vspace{-1.7ex}
\noindent where $c$,  % is a finite constraint in $\CStell$,
$c_{i}$ are finite constraints in $\CSdom$, $x \in \Var$,
$p\in\Pi$ (the set of all process symbols), $\vec{x}$ is a list of variable names corresponding to the formal parameters of process $p$,
and $n\in\N^{> 0}$.
A \tccp{} program
is a pair $\Qq{D}{A}$, where $A$ is the initial agent
and $D$ is a set of \emph{process declarations} of
the form $p(\vec{x}):- A$.

The \emph{operational semantics} of \tccp{}
is described by a transition system $T=(\Confc, \rightarrow)$.
Configurations in $\Confc$ are pairs $(A,c)$
representing the agent
$A$ to be executed in the current store $c$.
%
%In contrast to the \tccp{} approach, %in \hytccp{}
The  transition relation $\rightarrow \subseteq \Confc\times\Confc$
%does not represent the passage of one time unit.
%Instead, it
models a computational step which  consumes one step of discrete time
which is used to synchronize the agents in parallel.
% For this reason, the discrete transitions that consumes one
% time unit in \tccp{} becomes instantaneous transitions in \hytccp{}.
% The transition relation $\rad \subseteq \Confc\times\Confc$
% represents a \tccp{} discrete transition whose execution is
% instantaneous, while
%The continuous passage of time is modeled by the
%transition relation $\rac{\tau} \subseteq \Confc\times\Confc$ where
%$\tau \in \R^{>0}$ indicates the duration of the transition.
%
In Figure~\ref{fig:op_sem_tccp}, we formally
describe this operational semantics. %of \hytccp{}.
%Where possible, we use %the subindex
%$\lambda\in \R^{>0} \cup \{\dt\}$ to represent both kinds of transitions.
%(discrete and continuous).

\begin{figure}[t]
    \begin{minipage}{\linewidth}
          {\scriptsize
     {\setlength{\jot}{1ex}
        \begin{align*}
            % tell
            &\sequent{d  \neq \CSfalse}
                {(\atell{c},d)\rightarrow
                (\askip,c \CSmerge d)
                }{}{(\mathbf{tell})}%\label{rule:R1}
           \quad \quad \quad \quad
           \sequent{\exists\,1 \leq k \leq n\,.\, d \vdash c_{k}  \quad d \neq \CSfalse}
            {(\textstyle{\sum_{i=1}^{n}\aask{c_{i}} A_i},d)
             \rightarrow
             (A_k,d)
             }
            {}{(\mathbf{ask})}%\label{rule:R2}
            \\
            % now
            & \sequent{(A,d)
             \rightarrow
             (A',d') \quad d \vdash c  }
            {(\anow{c}{A}{B},d)
             \rightarrow (A',d')
             }
            {}{(\mathbf{now1})}%\label{rule:R3}
            \quad\quad
            \sequent{(A,d)
             \not \rightarrow  \quad d \vdash c  \quad d \neq \CSfalse}
            {(\anow{c}{A}{B},d)
             \rightarrow (A,d)
             }
            {}{(\mathbf{now2})}%\label{rule:R3'}
            \\
            & \sequent{(B,d)
             \rightarrow
             (B',d') \quad d \not \vdash c  }
            {(\anow{c}{A}{B},d)
             %\rightarrow (A,d')
             \rightarrow (B',d')
             }
            {}{(\mathbf{now3})}%\label{rule:R4}
            \quad\quad
            \sequent{(B,d)
             \not \rightarrow \quad d \not \vdash c  \quad d \neq \CSfalse}
            {(\anow{c}{A}{B},d)
             \rightarrow (B,d)
             }
            {}{(\mathbf{now4})}%\label{rule:R4'}
            \\
            % parallel
             & \sequent{ (A,d) \rightarrow(A',d') \quad
             (B,d) \rightarrow(B',d'')}{
             %\begin{array}{c}
                (A \parallel B,d) \rightarrow (A' \parallel B',d'\CSmerge d'') %\\
                 %\triple{B \parallel A}{d}{\tilde{d}} \rad \triple{B \parallel A}{d'\CSmerge d''}{\tilde{d}' \HCSmerge \tilde{d}''}
             %\end{array}
             }{}{(\mathbf{par1})}%\label{rule:R5}
            \quad\quad
           \sequent{ (A,d) \rightarrow (A',d') \quad
           (B,d) \not \rightarrow  }{
           \begin{array}{c}
               (A \parallel B,d) \rightarrow (A'\parallel B,d') \\
               (B \parallel A,d) \rightarrow (B \parallel A',d')
           \end{array}
           }{}{(\mathbf{par2})}%\label{rule:R8}
           \\
           & \sequent{ (A,l \CSmerge \CShid{x}{d})
          \rightarrow (B,l')}{
           (\ahiding[l]{x}{A},d)
           \rightarrow (\ahiding[l']{x}{B},d \CSmerge \CShid{x}{l'})}{
           }{}{(\mathbf{hid})}
           \quad\quad\quad \quad
           \sequent{ \mgrule{p}{x}{A} \in D  \quad d  \neq \CSfalse}
           { (\mgc{p}{x}{} ,d) \rightarrow (A,d)}{}{(\mathbf{proc})}
           \end{align*}
          }
        }
        \caption[The  transition system for \tccp{}.]{The transition system for \tccp{}.}
        \label{fig:op_sem_tccp}
    \end{minipage}
\end{figure}
Let us briefly describe the behaviour of each \tccp{} agent.
Agent $\askip$ ends the computation. Agent $\atell{c}$
adds $c\in\CSdom$ to  the store. Agent $\textstyle{\textstyle{\sum_{i=1}^{n}\aask{c_{i}} A_i }}$
allows the non-deterministic choice.
If a guard $c_i$ is entailed by the store, a  transition takes place and agent $A_i$ is executed (rule \textbf{ask}) in the next time instant.
% Finally, the passing of continuous time
% %has to be modeled.
% %To this aim we use
% has to be modeled by means of
% a new agent $\aaskc{\inv}$
% which makes continuous variables evolve
% over time while the invariant $\inv$ holds.
% The \tccp{} choice agent is extended
%If an invariant $\inv_j$ hold in the store
If no guard  is currently entailed by the global store, the choice agent suspends
and waits for one of its guards to be activated by a concurrent agent.

% In that case, the flow of the velocity $V$
% is changed to $10$.

The conditional agent $\anow{c}{A}{B}$
behaves as $A$ (\resp{} $B$) in case $c$
is (\resp{} is not) entailed by the store. It is worth noting that  \tccp\ handles negation as failure, this meaning that asking whether a constraint $c$ is held by the store $d$ produces false ($d \not \vdash c$) both when
$\neg c$ is entailed and when no information about $c$ can be deduced.
The parallel composition $\parallel$
is interpreted in terms of maximal parallelism, \ie{} at each step
all the parallel enabled agents can be executed simultaneously (rules \textbf{par1} and \textbf{par2}).

%\begin{pleasenote}
%La definicion de $\tilde{d} \lhd \tilde{d}'||\tilde{d}''$ y $\tilde{d}'-\tilde{d}$ es nueva. Trata de capturar la mezcla de dos stores continuous, teniendo en cuenta las posibles inconsistencias. En la regla par1, he cambiado $\tilde{d}'\tilde{\wedge}\tilde{d}''$ por este nuevo store.
%\end{pleasenote}

The hiding agent $\ahiding{x}{A}$ makes variable $x$ local to $A$.
Finally, $p(\vec{x})$ takes from $D$ a declaration of the form
$p(\vec{x}):-A$ and then executes $A$.

In order to
detect when the store becomes inconsistent
we explicitly check if $d\neq \CSfalse$ in the rules
in \smartref{fig:op_sem_tccp}.
This check follows the
philosophy defined for \ccp{} in \cite{SaraswatR90}
and for \tccp{} in \cite{CominiTV11absdiag}, where
computations that reach an inconsistent store are considered failed
computations.

Let us formalize the notion of behaviour of a \tccp{} program $P$
in terms of the transition system described in
\smartref{fig:op_sem_tccp}. The small-step operational behaviour of \tccp{}
collects all the small-step computations
associated with $P$
%as the set of all (the prefixes of) the sequences of
%stores
(in terms of sequences of \tccp{} stores closed by prefix)
for each possible
initial store.

% \footnote{We assume that our system does
% not exhibit Zeno behaviors.}.

% Using the transition system described in \smartref{fig:op_sem}, we
% define the notion of behavior ($\Beha{D}$) that, intuitively,
% collects all the small-step computations associated to a set of
% declarations $D$ as the set of (all the prefixes of) the sequences
% of computation steps for all possible initial agents and stores.
%
%\begin{pleasenote}[Laura]
%El titulo de la definicion ocupa el margen. No se como modificar el entorno para que este el titulo en dos lineas
%\end{pleasenote}
%\begin{pleasenote}[LLM]
%Arreglado :)
%\end{pleasenote}
\begin{definition}[\small{Small-step observable behaviour of \tccp}]
    \label{def:ssBeha}
    Let $P=D.A$ be a \tccp{} program.
    The \emph{small-step (observable) behaviour} of $P$ is defined as:
    \begin{align*}
        % \Beha{D} &\dfn \bigcup_{\forall \hst{c}{\tilde{c}}\in\CSdom\times\HCSdom, \forall A\in\Agents}
        % \BehaQ{\hst{c}{\tilde{c}}}{A}{D} \qquad \text{where} \\
        \mathcal{B}^{\mathit{ss}}\lsem D.A \rsem&\dfn \bigcup_{c_0\in\CSdom} \big\{
        c_0 \cdot  c_1
        \cdot \ldots \cdot c_n
        \mid \\[-2ex]
        &\qquad \quad (A,c_0) \rightarrow
        (A_1,c_1)\rightarrow \dots \rightarrow
       (A_n,c_n)
        \big\} %\cup \{\ecrs\}
    \end{align*}
    where $\rightarrow$
    is the transition relation given in \smartref{fig:op_sem_tccp}.
    %
    % We call the sequences in $\BehaQ{c}{A}{\P}$ \emph{behavioral timed
    % traces} or simply \emph{traces} (when clear from the context).
    %
    % We denote by $\BehaEq{}{}$ the equivalence relation between process
    % declarations induced by $\Beha{}$, namely for all
    % $\P[1],\P[2]\in\Progs$, $\BehaEq{\P[1]}{\P[2]} \iff \Beha{\P[1]} =
    % \Beha{\P[2]}$.
\end{definition}
%\textbf{Abstract Machine}

\subsection{Example of \tccp}\label{sec:photocopier_example}

We show an example of a \tccp{} program in
Figure~\ref{fig:tccp-photocopier}. This program (extracted from~\cite{Alpuente200558}) models a photocopier
by means of four procedure declarations which represent the  user
process  \texttt{user(C,A)}, the photocopier
\texttt{photocopier(C,A,MIdle,E,T)}, the system process \texttt{system(MIdle,E,C,A,T)}  and the intialization of such
processes
\texttt{initialize(MIdle)}.

\begin{figure}[tb]
 {\scriptsize
\begin{center}
\begin{minipage}{10cm}
\begin{tabbing}
{\tiny 1} \quad\tt user(C,A):-  \=\tt \tt \textsf{ask}(\lista{A}{free}{\_}) $\rightarrow$ (\textsf{tell}(\lista{C}{on}{\_})+ \textsf{ask}(\lista{A}{free}{\_}) $\rightarrow$ (\textsf{tell}(\lista{C}{off}{\_})+ \\
{\tiny 2}\tt\tt 	   \>	\tt\textsf{ask}(\lista{A}{free}{\_}) $\rightarrow$ (\textsf{tell}(\lista{C}{c}{\_})+ \textsf{ask}(\lista{A}{free}{\_}) $\rightarrow$ (\textsf{tell}(true)).
\end{tabbing}
\begin{tabbing}
{\tiny 3} \quad\tt pho\=\tt tocopier\=\tt(C,A,MIdle,E,T):- $\exists$ \tt Aux,Aux',T'(\textsf{tell}(\lista{T}{Aux}{T'})|| \\
{\tiny 4}	\tt\tt 	   \>	\textsf{ask}\=(true) $\rightarrow$ \=\tt \textsf{now} \=\tt (Aux > 0) \textsf{then} \\
{\tiny 5}			\>		\>		\>		\>\tt 	\textsf{now} \=\tt (\lista{C}{on}{\_}) \textsf{then}  \textsf{tell}(\lista{E}{going}{\_}) ||  \textsf{tell}(\lista{T'}{MIdle}{\_})|| \textsf{tell}(\lista{A}{free}{\_}) \\
{\tiny 6}			\>		\> 		\>		\>\tt	\textsf{else}  \=\tt \textsf{now} \=\tt (\lista{C}{off}{\_}) \textsf{then} \textsf{tell}(\lista{E}{stop}{\_}) ||  \textsf{tell}(\lista{T'}{MIdle}{\_})|| \textsf{tell}(\lista{A}{free}{\_}) \\
{\tiny 7}			\>		\>		\>		\>		\>\tt \textsf{else} \=\tt \textsf{now} \=\tt (\lista{C}{c}{\_}) \textsf{then} \textsf{tell}(\lista{E}{going}{\_}) || \textsf{tell}(\lista{T'}{MIdle}{\_})|| \textsf{tell}(\lista{A}{free}{\_})\\	
{\tiny 8}			\>		\>		\>		\>		\>			\>\tt \textsf{else} \=\tt \textsf{tell}(Aux'=Aux-1) || \textsf{tell}(\lista{T'}{Aux'}{\_}) || \textsf{tell}(\lista{A}{free}{\_})\\
{\tiny 9}			\>		\>		\>\tt 	\textsf{else} \textsf{tell}(\lista{E}{stop}{\_}) || \textsf{tell}(\lista{A}{free}{\_})).  					
\end{tabbing}
\begin{tabbing}
{\tiny 10} \quad\tt system \=\tt (MIdle,E,C,A,T):- $\exists$ \tt E',C',A',T'\\
{\tiny 11}		\>\tt	(\textsf{tell}(\lista{E}{\_}{E'}) || \textsf{tell}(\lista{C}{\_}{C'})|| \textsf{tell}(\lista{A}{\_}{A'}) || \textsf{tell}(\lista{T}{\_}{T'}) || user(C,A) ||\\
{\tiny 12}        \>		  \tt   \tt \textsf{ask}(true) $\rightarrow$ (photocopier(C,A',MIdle,T,E')) || \textsf{ask}(\lista{A'}{free}{\_}) $\rightarrow$ (system(MIdle,E',C',A',T'))).
\end{tabbing}
\begin{tabbing}
{\tiny 13} \quad\tt initialize\=\tt(MIdle):-\,$\exists$ E,C,A,T	\\
  \>		  \tt   \tt (\textsf{tell}(\lista{A}{free}{\_}) || \textsf{tell}(\lista{T}{MIdle}{\_})||  \textsf{tell}(\lista{E}{off}{\_}) || system(MIdle,E,C,A,T)). 		
\end{tabbing}
\end{minipage}
%}
\vspace{-1ex}
\caption{A \tccp{} program modelling a photocopier}
\label{fig:tccp-photocopier}
\end{center}}
\vspace{-.5ex}
\end{figure}

Streams \texttt{C} and \texttt{A}  are the communication channels through which the
 \texttt{user} sends commands to  the \texttt{photocopier}, and the \texttt{photocopier} communicates
 its state to the \texttt{user}, respectively.  The \texttt{user} waits for the \texttt{photocopier} to be \texttt{free} to send it a new command (make a copy (\texttt{c}), turn on/off (\texttt{on}/\texttt{off}) or do nothing (\texttt{true})), which is non deterministically chosen. Agent \texttt{photocopier} uses stream \texttt{T} as a counter to check whether a command has been received during \texttt{MIdle} time units and, in another case, to automatically turn off.
 In the case deadline \texttt{MIdle} has not been reached, agent \texttt{photocopier} accepts the command sent by the \texttt{user}, and behaves accordingly, updating its local state, in stream \texttt{E}, and counter \texttt{T}.

Agent \texttt{system} is in charge of creating and synchronising agents \texttt{photocopier} and \texttt{user} correctly. Finally, \texttt{initialize} creates the initial agents and streams and establishes the value of the time deadline \texttt{MIdle}.

The example shows several characteristics of \tccp. For instance, the  use of streams as $\ask$ guards in agent \texttt{user} (lines 1-2) is a usual way of modelling agent synchronisation and communication. In this case, we guarantee that the \texttt{user} does not send a new command until the \texttt{photocopier} has processed the previous one and, therefore, it has instantiated the head of stream \texttt{A} to \texttt{free} (rule \textbf{ask} of Figure~\ref{fig:op_sem_tccp}). In addition, the combination of \texttt{tell} and streams makes it possible to extract values from the store. For example, agent \texttt{tell(T = [Aux|T'])} (line 3) in \texttt{photocopier} adds constraint \texttt{T = [Aux|T']} to the store (rule \textbf{tell} of Figure~\ref{fig:op_sem_tccp}). But, if a constraint such as \texttt{T = [v|\_ ]} already exists in the store, the agent has the side effect of binding \texttt{Aux} to \texttt{v} and the tail of the stream to \texttt{T'}.

It is worth noting the use of agent \texttt{now} in agent \texttt{photocopier}. As rules for \textbf{now} in Figure~\ref{fig:op_sem_tccp} show, agent \texttt{now} is able to handle both positive and negative information from the store. Thus, for instance, in lines 4-8,  \texttt{photocopier} reads and processes the command sent by the \texttt{user}, but if no command has been delivered (the store has no information about variable \texttt{C}), agent  \texttt{now}  proceeds in the \texttt{else} branch (line 9).

The store may hold different types of constraints. In addition to the logic constraints on streams, the example contains linear constraints on numerical variables which can be added/read to/from the store (lines 4 and 8).

The example also shows the extensive creation, through agent $\exists$, of new local variables (rule \texttt{hid} of Figure~\ref{fig:op_sem_tccp}). When streams are used as communication channels between agents, the proliferation of variables in \tccp\ is a consequence of the monotonic character of the store. Specifically,  the evolution over time of agents \texttt{user} and \texttt{photocopier} involves the addition of new values to the  streams for which new references to the stream tail are needed (lines 3 and 10). Observe that to simplify the \tccp\ syntax nested $\exists$ agents are collapsed, that is, for instance, $\exists$ \texttt{Aux}($\exists$ \texttt{Aux'}($\exists$ \texttt{T'} $\cdots$)) is written as $\exists$ \texttt{Aux,Aux',T'}($\cdots$).

Finally, note that rules \textbf{par1} and \textbf{par2} of  Figure~\ref{fig:op_sem_tccp} synchronise \tccp\ agents completely, that is, at each time instant, all agents that can evolve, proceed synchronously. This is why sometimes it is necessary to include delays (\texttt{ask(true)}) in the agents. For instance, the delay in line 4 of  \texttt{photocopier} guarantees that when the  following agent \texttt{now} is executed, the value of variable \texttt{Aux} has been correctly extracted from the store in line 3.

\section{Architecture of the proposal}\label{sec:approach}
The development of tools for \tccp\ holds many challenges, due to its logic, concurrent and synchronous nature.
The main problems we face are:
\begin{itemize}
\item Dynamic generation of fine-grained procedures: \tccp\ procedures usually have a short live cycle, but with a cyclic or recursive behaviour. The correct identification of the procedures instances and their scope is a challenge when implementing tools.
\item Dynamic generation of (local) variables: the dynamic generation of procedures creates a large number of variables, local to each procedure instance. In addition, the typical syntax of streams also abuses fresh variables. The challenge is how to correctly identify variables and store them.
\item Parallel execution of agents: \tccp\ agents executing in parallel can have a different view of the store, although it is a shared and single memory. It is a challenge to keep the different views consistently.
\item Constraint solving: the dynamic generation of procedures and variables makes it challenging to solve constraints efficiently.
\end{itemize}

We propose an architecture that facilitates the development of tools for the simulation and analysis of \tccp, addressing the aforementioned problems modularly
which gives us independency from the final implementation platform.
%In this section, we present an architecture that simplifies the development of tools, which assists in the analysis of concurrent
%systems described by languages like \tccp.
Our proposal is based on the definition of the so-called \tccp\ abstract machine  where the different elements of the \tccp\ semantics are separated following a structured methodology which leads to the construction of modular, extensible and reusable tools. Although we are aware that the machine is not abstract in the classic sense, we call it so because it constitutes the execution core of the abstract instructions.

% which allows us to separate the semantics
%of the language from the specific implementation.

\begin{figure}[t]
	\centering
 		\includegraphics[width=0.45\textwidth]{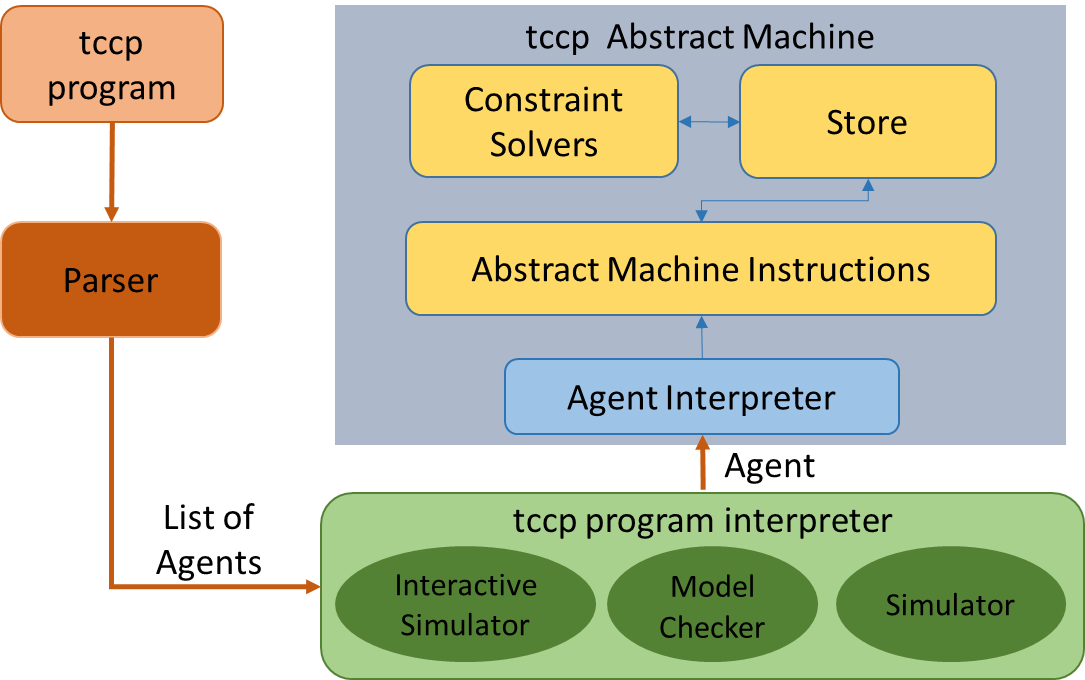}
	\caption{The architecture of the proposal}
	\label{fig:architecture}
\vspace{-.5ex}
\end{figure}

\smartref{fig:architecture} depicts the architecture of our proposal.
%, which can be used to develop tools like simulators, model checkers, etc.
The input of any tool is a \tccp\ program in plain text that is transformed into an intermediate code that the \tccp\ machine can interpret.
The \texttt{program interpreter}  takes the transformed \tccp\ program and controls its execution  at a high level, i.e., it commands the \tccp\ machine to run the different program agents, but it is unaware of the semantics of the agents.
The \texttt{program interpreter} may be implemented as a simulator, an interactive simulator, or even as a model checker if the memory structures to record the whole search state space are added.

The \texttt{abstract machine} defines a set of  \texttt{instructions} that model the basic operations on the global store. The behaviour of the different \tccp\ agents is implemented using these instructions, which bridge the gap between the modelling language and the language used in the implementation.
These basic instructions work on an abstract view of the \texttt{store}, i.e., they do not take into account the type of constraints handled by the language,
or how memory is really managed, which will depend on each specific implementation.

The abstract machine includes an \texttt{agent interpreter} that knows the behaviour of each agent, given as a sequence of basic instructions.
%These functions interact with the store to add constraints  to its current state, without knowing how the memory model is implemented and constraints are managed.
Finally, the \texttt{store} module has a dual role. On the one hand, as explained above, it provides an interface to the abstract machine that has an abstract view of actual memory model. On the other hand, because \tccp\ is a constraint-based paradigm, the store makes use of a constraint solver to manage the constraint operations.

This kind of architecture is highly modular, which has several advantages. If the semantics of an agent is modified, we only have to change the agent interpreter. If the \tccp\ language is extended with new agents, we also have to slightly modify the parser. Similarly, if we wish to support other types of constraints, or to solve constraints more efficiently, we only have to revise the constraint solver and the implementation of the store.

We now describe the main elements of this proposal in more detail: the store, the instructions of the abstract machine and the agent interpreter. 
\subsection{Store}
\label{sec:store}

The store is the {\em memory} of the abstract machine, and is in charge of keeping the constraints over variables. The store is unique during the execution of \tccp\ programs, which means that all agents access  the same memory structure. In \tccp, the preservation of the store consistency among all agents in execution is essential  to guarantee the correct implementation of the parallel operator (rules \textbf{par1} and \textbf{par2} of \smartref{fig:op_sem_tccp}).

An agent of \tccp\ is a light process with a short execution time.
The natural mechanism of execution involves the creation of a large number of variables,
with a restricted scope. We solve this issue by dividing the store into two memory elements: the symbol table
and the global memory.

%\paragraph{Symbol table:\\}
The symbol table is a tree structure used to determine the scope of a variable.
Each tree node contains the local variables visible for a set of agents.
There are two \tccp\ agents that can define new variable scopes and, therefore, can add new nodes to the tree: exists and procedure call agents (rules \textbf{hid} and \textbf{proc} of \smartref{fig:op_sem_tccp}).
The scope of the rest of the agents is associated with an already existing node.
%\paragraph{Global memory:\\}
The global memory is responsible for keeping
the constraints over the variables and provides the consistency of the store.

\subsubsection{Abstract Machine Instructions}\label{sec:basic_functions}

We now enumerate the set of basic instructions of the abstract machine  used to implement the execution of agents. We need some preceding definitions.
Let  $x \in Var$, and $A$ be a \tccp\ agent. %For the sake of simplifying the notation, we allow $A$ to be a choice branch such as $\aask{c} A'$.
%In the sequel, we use the following notation.
\begin{itemize}
%	\item $A.node$ denotes the node of the symbol table which determines the scope of the agent $A$.
%	\item $A.cnst$ denotes constraint associated with  $A$, that is, if $A = \tell(c)$, $A = \aask{c} A'$ or $A = \now(c) \cdots$ then $A.cnst = c$. Otherwise, that is, if $A$ is any other \tccp\ agent, $A.cnst = \true$.
%	\item $A.cnst.vars$ is the set of referenced variables in constraint $A.cnst$,  all of them defined in the scope $A.node$.
	\item $A.x$ denotes the variable named $x$ in the scope of $A$.
\item $p.\vec{x}$ represents the formal parameter $\vec{x}$ of procedure $p$ in the \tccp\ program.
	%\item $A.agents$ is the set of agents that can be executed
	%after running $A$, that is, in the next execution step. For instance,
%\begin{itemize}
%\item if the current store entails $c$,  then $(\aask{c} A').agents = \{A'\}$, otherwise,  $(\aask{c} A').agents = \{\aask{c} A'\}$;
%\item if $A = \tell(c)$ then $A.agents = \emptyset$;
%\item  if the current store entails $c_1$ and $c_2$, then $(\aask{c_1} A_1 || \aask{c_2} A_2).agents = \{A_1,A_2\}$, and so on.
%\end{itemize}
	%	in the next step after the $A$ execution.
\end{itemize}

In the description below,  $global$ is the global store of
the abstract machine (comprising the symbol table and the global memory),
 $A$ is the current agent to be executed by the machine, $c$ is a constraint, and  $local_1$ and $local_2$ are the local stores produced by the parallel execution of agents.

The abstract machine provides the following basic functions:

\begin{itemize}
	\item $is\_consistent():Boolean$: returns whether or not $global$ is consistent. %The return parameter is a boolean.
	%\item $is\_defined(A.x):Boolean$: checks if variable $x$ of agent $A$ exists in the store. This method can be extended for a list of variables.
	\item $add\_variable(A.x)$: adds  variables $x$ to $global$. Variables $x$ must be local to agent $A$.
	\item $add\_parameter(p.\vec{x},\vec{x'})$: given a call $p(\vec{x'})$ to a \tccp\ procedure $p(\vec{x})$, adds new variables $\vec{x}$ to $global$ and links it to the variables $\vec{x'}$ used in the procedure call.
	\item $add\_constraint(c)$: adds constraint $c$ to $global$.
	\item $entails(c):Boolean$: checks if  $global$ entails constraint $c$.
   \item $merge(local_1,local_2)$: updates  $global$ with the new constraints added to $local_1$ and $local_2$, if it is possible, that is, if no inconsistencies exist between the constraints in $local_1$ and $local_2$.
   %\item $upadate(local)$: updates  $global$ with the new constraints added to $local$.
\end{itemize}

\subsubsection{Agent Interpreter}\label{sec:agent interpreter}

\begin{figure}[t]
	\centering
 		\includegraphics[width=0.50\textwidth]{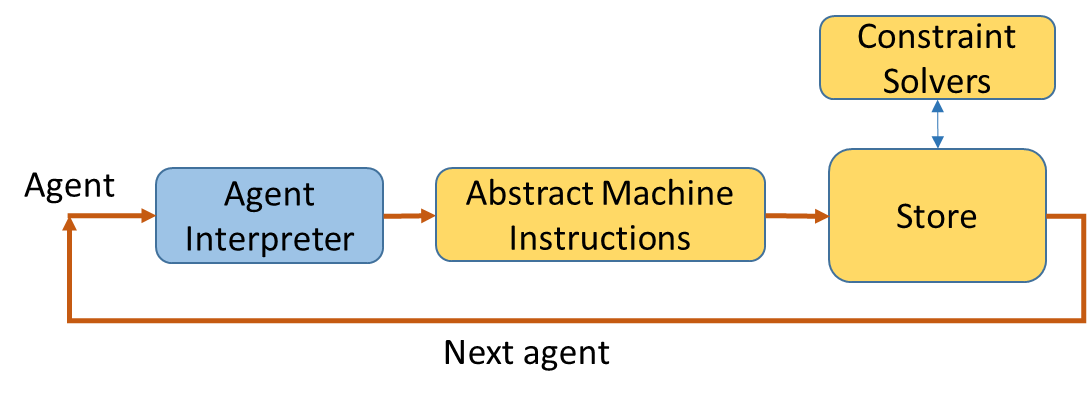}
	\caption{Interpreter of agents}
	\label{fig:agentInterpreter}
\end{figure}

The agent interpreter transforms each agent into the sequence of abstract machine instructions following the semantics given in \smartref{fig:op_sem_tccp}.
\smartref{fig:agentInterpreter} depicts how the agent interpreter works. Given the current $A$, the interpreter first executes the corresponding abstract machine instructions,
which directly interact with the store $global$, and then determines the  agent $nA$ to be next executed.

In the following paragraphs, we define the execution of the current agent $A$, denoted as {\em execute(A)},
inductively on the syntactic structure of $A$. We assume that the agent $A$ is executed on the global store $global$. Function $execute(A)$ returns the new store $local$ produced by the execution of agent $A$ and the set of agents to be next executed.  %We assume that the execution of an agent returns a list
%of agents to be executed next,  we denote this list as.
Some agents, such as \askip\ and \tell, always return an empty set of agents.
%Given a store d, the behaviour of each agent is defined below:

\paragraph{parallel:} $A_1 || A_2$
%$A_1 || A_2$ is executed as follows:
\begin{enumerate}
	\item $is\_consistent()$: if it is true, continue to step 2 else stop
	\item Let $execute(A_1)=\langle local_1, nA_1\rangle$
    \item Let $execute(A_2)=\langle local_2, nA_2\rangle$
	\item Return $\langle merge(local_1,local_2), nA_1 || nA_2\rangle$
\end{enumerate}

\paragraph{tell(c)\\}
%Given a store $d$, the behaviour of $\atell{c}$ is defined as follows:
\begin{enumerate}\vspace{-0.3cm}
	\item $is\_consistent()$: if it is  true, continue to step 2 else  stop
	%\item $d.is\_defined(tell.c.vars)$: if all variables are defined in $d$, proceed with step 3. If not, set $d$ as inconsistent.
	\item $add\_constraint(c)$
	\item Return $\langle global, \askip \rangle$
\end{enumerate}

\paragraph{choice:} $\sum_{i=1}^{n}\aask{c_{i}} A_i$

%An agent $\aask{}$ is part of a choice agent of the form.
%Given a store $d$, the execution of a generic choice is as follows:
\begin{enumerate}
	\item $is\_consistent()$: if it is  true, continue to step 2 else stop
	\item If $\neg entails(c_i)$ for $i= 1, \dots ,n$,  proceed with step 4, else select randomly one branch $\aask{c_{i}} A_i$ such that $entails(c_i)$ holds and proceed with step 3
	%For each  $i= 1, \dots ,n$ the checking process consists of:
%	\begin{enumerate}
%		\item $d.is\_defined(choice.c_i.vars)$: if all variables are defined in $d$, proceed with the next step. If not, check $c_{i+1}$.
%		\item $d.entails(choice.c_i)$: if $d$ entails the constraint $c_i$, add $A_i$ to the list of possible executed agents.
%	\end{enumerate}	
	\item Return $\langle global, A_i \rangle$
	\item Return  $\langle global,\sum_{i=1}^{n}\aask{c_{i}} A_i \rangle$
\end{enumerate}

\paragraph{now:} $\anow{c}{A}{B}$

%Given a store $d$, the execution of  $\anow{c}{A}{B}$ is defined as follows:
\begin{enumerate}%\vspace{-0.3cm}
	\item $is\_consistent()$: if it is true, continue to step 2 else stop
%	\item $d.is\_defined(now.c.gvars)$: if all variables referenced in $c$ are defined in $d$ we can proceed with step 3. If not, set $d$ inconsistent.
	\item if $entails(c)$ then return $execute(A)$, else return $execute(B)$
	%%\item if branch {\em then} was executed, return $nextAgents = execute(A)$. If {\em else} branch was executed, return $nextAgents = execute(B)$.
	%\item return $nextAgents = execute(A)$ or $nextAgents = execute(A)$ according to the branch executed.
\end{enumerate}

\paragraph{hiding:} $\ahiding{x}{}$\\
%Given a store $d$ and a variable $X$, the behaviour of $\ahiding{x}{A}$ is defined as follows:

\begin{enumerate}\vspace{-0.3cm}
	\item $is\_consistent()$: if it is  true, continue to step 2, else stop
	\item $add\_variable(A.x)$
	\item Return $execute(A)$ %update the list of agents to be executed in the next step $exists.agents.add(A.agents)$.
\end{enumerate}

\paragraph{procedure call:}$p(\vec{x}):- A$\\
%Given a store $d$ and a program $p$ with parameter $X$, the behaviour of $p(X):- A$ is defined as follows:

\begin{enumerate}\vspace{-0.3cm}
	\item $is\_consistent()$: if it is  true, continue to step 2 else stop
	\item $add\_parameter(p.\vec{x},p.\vec{x'})$, where $\vec{x'}$ are the variable used in the procedure call $p(\vec{x'})$
	\item Return $\langle global,A \rangle$ %Add agent $A$ to the list of agents to be executed in the next step, $p.agents.add(A)$.
\end{enumerate}

Observe that the implementation of agents described above closely follows the agent semantics of \smartref{fig:op_sem_tccp}. Thus, proving the correctness of the implementation reduces to proving that the basic instructions of Section~\ref{sec:basic_functions} update the store correctly.

%\begin{pleasenote}[MMar]si me diera tiempo, podria incluir algo sobre la correccion de la implementacion
%\end{pleasenote}

\section{Implementation issues}
\label{sec:implementation}
We have implemented a \tccp\ simulator based on the abstract machine presented in Section~\ref{sec:approach}.
The  tool accepts \tccp\ programs with linear constraints over arithmetic variables, and logic constraints over streams.
A stream can store text or linear expressions over arithmetic variables.

%The tool implements a discretised version of the semantics, which uses convex polyhedra to abstract continuous behaviour.
%For this reason, the value and flow of continuous variables is defined with rectangular sets, and the tool supports linear constraints over continuous variables.
%The simulator uses \ppl\ to represent and manage convex polyhedron.

%\begin{pleasenote}[MMar]yo creo que a la parma poliedra habria que referenciarla, y explicarla un poco...
%\end{pleasenote}
%\begin{pleasenote}[Laura] He añadido una referencia y poco más sobre la ppl
%\end{pleasenote}
The \tccp\ machine has been implemented in \java. We have used existing third party \java\ libraries to implement the different elements of the simulation tool.
The \tccp\ machine includes two different constraint solvers, one for linear constraints and another for logic constraints.
The first one is based on Parma Polyhedra Library~\cite{BagnaraHRZ05} (\ppl), a library that provides numerical abstractions such as convex polyhedra or grids. The constraint solver for linear constraints uses convex polyhedron to represent a system of linear constraints, and each dimension represents a variable. The polyhedron is {\em universe} (true) when all variables are unconstrained, and it is empty (false) when constraints are not satisfied. Moreover, \ppl\ also provides methods
to check if the polyhedron entails a specific constraint. The logic constraint solver has been developed from scratch,
since it strongly depends on the memory model implemented.
%This solver provides methods to walk through the data stored in a stream variable, and compare its content.
There are other \java\ constraint solvers, such as JaCoP~\cite{jacop_web}, which supports a large variety of constraints (basic arithmetic operation constraints, logical and conditional constraints, regular constraints for the assignment to variables, etc.). This constraint solver is versatile and easy to use in solutions such as this proposal. However, we have preferred to use \ppl, since we plan to extend our abstract machine and the simulation tool for modelling rectangular hybrid systems and \ppl\ provides more suitable numeric abstraction to represent continuous variables and their evolution over time.

%% Parser
We have used ANTLR~\cite{Parr2013} to generate the parsers included in the simulator.
Given a grammar described in the notation similar to BNF, ANTLR produces the base classes for the parser and visitor.
We have extended the base visitor to control how to walk the parse-tree, and to specify the returned type.
The tool includes a parser that transforms a \tccp\ program from plain text into a list of \texttt{Agent} objects.
In addition, there are independent parsers for linear constraints and logic constraints.
They return respectively, \ppl\ Constraint System and Stream objects.

\subsection{Store implementation}
\label{sec:storeimplementation}
%\subsubsection{Hy-tccp Store}\label{subsec:store}

%The discrete store used by \hytccp, which is inherited from tccp, can be viewed as a blackboard where information is continuously written and never cancelled. The architecture of the monotonic store for our implementation is made up of two structures: the table of symbols and the global memory. The first one store the variables defined in the \hytccp\ program. The global memory stores the current value/content of the variables. Both structures are monotonic in the sense that once a element is added, it cannot be removed.
%
%As discussed earlier in Section \ref{subsection:abstMachine}, the decision of the type of structure of the table of symbols was taken by the hiding mechanism of \hytccp, used by agents $\ahiding{x}{A}$ that create new variables that are only available in the scope of A, and by the procedure calls. We needed a structure that makes invisible some variables and don't allow the access to elements out of its scope. For this reason, we have designed the table of symbol as a tree of table of symbols. Each single node of the tree is composed of an array of tuples. Three elements define a tuple in the table of symbols: the name, the type and a reference to the global memory element that stores the content of the variable. To insert a new variable in the table of symbols, it is necessary to know its scope to add it in the correct node of the tree. Once we have the node, the scope of the element, this is added at the end of the list.

%---- Laura ---
The store, presented in Section~\ref{sec:approach},
%, is divided in the symbol table and the global memory. This memory model
has to keep and manage logic constraints over streams, and linear constraints over numeric variables, which are represented by convex polyhedron.
To this end, the memory model is extended with a convex polyhedron, called $disc\_poly$,
that saves linear constraints.
Each dimension of  $disc\_poly$ represents a variable. The dimension assigned will be the same until the end of the program execution.
Consistently with the notion of store in \tccp, polyhedron $disc\_poly$ is monotonic, that is, constraints are added but never removed.
%Since continuous variables can be reset, $cont\_poly$ is not monotonic, and constraints can be added and removed from it.
%There is also an auxiliary polyhedron that stores the flows associated to each continuous variable.

The symbol table is a tree, whose nodes have an identifier and store the list of variables belonging to this scope.
For each variable, the node keeps the symbol identifier and the memory position that saves its information.
The need to use a tree instead of a list will be clear later, when we discuss the dynamic procedure generation.
%Its implementation is based on a tree of nodes as explained in Section~\ref{sec:abstract_store}.

The global memory is an array of registers, which keep the type of memory element, and a data field.
Currently, there are four types of memory elements, and each type saves different information in the data field:
\begin{itemize}
\item Constant: the data field stores the value of the constant.
\item Discrete variable: the data field stores the dimension that represents this variable in $disc\_poly$.
%\item Continuous variable: the data field stores the dimension that represents this variable in $cont\_poly$.
\item Reference:
%the element is a reference to other memory position.
the data field saves the referenced memory position.
\item Functor: the element is a stream with head and tail. The data field keeps the memory position of the head.
The tail is always stored in the next position from the head.
\end{itemize}

Below, we address the main issues of the store implementation, most of them are related to the characteristics of \tccp\ enumerated in Section~\ref{sec:approach}.

\paragraph{Store consistency\\}
The store includes diverse structures that keep the different kinds of constraints and variables. The global consistency or inconsistency
is determined as follows: the store is consistent if constraints over streams are consistent and $disc\_poly$ is not empty. In any other case, the store is inconsistent.

\paragraph{Dynamic procedure generation\\}
Every procedure
%A procedure call agent defines a new scope. Thus, every
call adds a new node  to the symbol table, and links it with its father node.
The identifier of the new node will be propagated, if necessary,  to the nodes of the agents in its scope.
%and propagates this node  to the agents included in its body.
This new node contains the list of its parameters. If a parameter is associated with a variable or constant value, it points to the caller variable/constant position. If a parameter is associated with an arithmetic expression, the parameter is linked to a variable that is constrained by this expression. Observe that when more than one procedure calls are executed in parallel, the new nodes created have the same father node, and this is why the symbol table built is a tree.

\paragraph{Dynamic variable generation\\}
The execution of an {\em exists} agent
adds
%creates a new scope; that is,
a new node  in the symbol table that identifies the scope of the variables.
%If there is a variable with the same name in an external scope, an agent included in the parentheses cannot see it.
The symbol table identifies and manages the potentially large number of variables with repeated names.
Similarly to the dynamic procedure generation, the identifier of the {\em exists} node will be propagated, if necessary,  to the nodes of the agents in its scope.
When these agents are executed, they start looking for symbols in the node of its exists parent. If the symbol is not found, they look in the parent of the node, and repeat this process until they find the symbol or reach a node created by a procedure call.

\paragraph{Concurrency in the store\\}
%An agent of \tccp\ executes concurrently with probably a different view of the store.
Agents executing in parallel have probably  different views of the store, which can involve consistency problems.
%without interfering with each others executions.
%In addition, each agent can have a different scope.
We address this issue by executing each agent with a copy of the memory structures that store the constraints. When concurrent agents have been executed, all the copies  are merged in such a way that the resulting store is in the right state (consistent or not) and includes the right constraints. For instance, if $agent_1$ and $agent_2$ add constraints over arithmetic variables, each agent will modify a copy of $disc\_poly$, named $p_1$ and $p_2$, and after executing both agents $disc\_poly = p_1 \cap p_2$.

\section{Evaluation}
\label{sec:evaluation}
In this section, we evaluate the current simulator running the {\em photocopier} example, presented in \smartref{sec:photocopier_example}. This example has an infinite and non-deterministic behaviour. In the evaluation, we will execute a finite number of steps of the abstract machine. In addition, we need to repeat the same trace several times to obtain statistics. To this end, the \texttt{user} process always selects the last branch of the choice; that is, the \texttt{user} does not send commands to the \texttt{photocopier}.

\begin{figure*}[t]
\begin{center}
\includegraphics[width=0.6\textwidth]{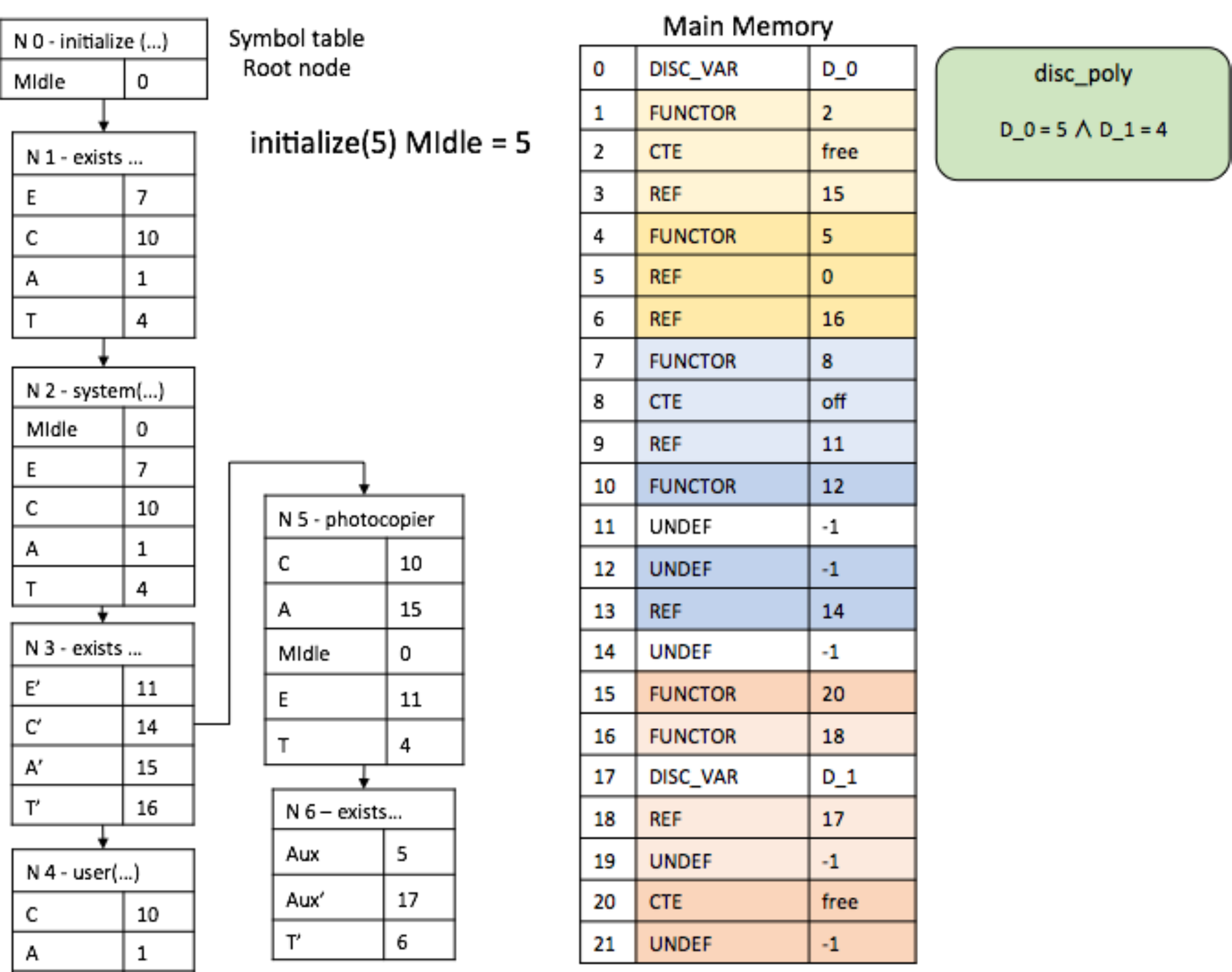}
\caption{Photocopier example - store state 7 steps}\label{fig:copier_execution}
\end{center}
\end{figure*}

%\begin{figure*}[t]
%\begin{center}
%\includegraphics[width=0.6\textwidth]{./figures/trainExecution_10steps.pdf}
%\caption{Railway example - store state 10 steps}\label{fig:train_execution}
%\end{center}
%\end{figure*}

We execute the example using the call \texttt{initialize(MIdle) || \tell(MIdle = 5)}.
\smartref{fig:copier_execution} depicts the state of the store after carrying out 7 steps of the simulation.
On the left hand side, we have the symbol table in which each node is created by a procedure call or an \texttt{exists} agent.
The root node (N0) is created because the procedure call \texttt{initialize(MIdle)} that contains the parameter \texttt{MIdle} set to 5.
In the next step, an \texttt{exists} agent is executed which creates the node N1 containing the list of local variables.
Then, several $\tell$ agents and a procedure call to \texttt{system} are executed in parallel. The $\tell$ agents add values to the heads of \texttt{A}, \texttt{T} and \texttt{E}, and the procedure call adds a new node (N2) with the list of parameters of \texttt{system} pointing to the variables used in the call.
%Now, a parallel agent is executed, so the variables A, T and E updates and then, we have to run the procedure call of system which implies to create a new node(N2), to store and reference the parameters of this.
In the following step, the \texttt{exists} agent is executed which creates a new node (N3),
and the parallel agent comprising four $\tell$ agents, a procedure call to \texttt{user} and two $\ask$ agents. The first $\ask$ can proceed, but the second one blocks because \texttt{A'} is unbound. Due to parallel execution of the procedure calls \texttt{(user(C,A)} $||$ \texttt{photocopier(C,A',MIdle,E',T'))}, nodes N3 and N4 are created. The rest of the nodes are generated in a similar way.

The main memory, on the right, holds information about the value of the variables during the execution. For example, positions 0 and 17 store the information of the  variables of the system (\texttt{MIdle}, \texttt{Aux} and \texttt{Aux'}). Observe that \texttt{Aux} (position 5) is a reference to position 0, which keeps a  variable represented in $disc\_poly$ by dimension 0. Another type of element in memory are functors, which represent stream variables. For example, position 1 is a functor whose head is a constant set to \texttt{free}, and whose tail is a reference to 15.

\begin{table}[t]
\caption{Evaluation results}
\begin{center}
\begin{tabular}{|l|r|r|r|}
\hline

\hline  & \textbf{30 steps} & \textbf{100 steps} & \textbf{500 steps} \\
\hline Symbol Table (nodes)& 26 & 85 & 417\\
\hline Global Memory  (regiters) & 78 & 239 & 1169\\
\hline disc\_poly  (dimensions) & 6 & 6 & 6\\
\hline 
\hline Heap used (MB)& 4 & 4.1 & 7 \\
\hline Heap allocated (MB)& 16.3 & 16.3 & 16.3 \\
\hline Parser (ms) & 192 & 196 & 197\\
\hline Simulation (ms)& 87& 192 & 2,170\\
\hline
\end{tabular}
%\begin{tabular}{l r r}
%& & \\
% SymbolTable (nodes). & GlobalMemory (regiters). & disc\_poly(dimensions).
%\end{tabular}

%\begin{tablenotes}
%{\scriptsize
%\item SymbolTable (nodes). GlobalMemory (regiters). disc\_poly(dimensions).
%}
%\end{tablenotes}
\end{center}
\label{tab:structures}
\vspace{-1.5ex}
\end{table}

%\begin{subtable}{.5\textwidth}
%\caption{Profiler statistics}
%\begin{center}
%\begin{tabular}{|l|r|r|r|r|}
%\hline
%  & \multicolumn{2}{c}{photocopier} \vline \\%& \multicolumn{2}{c}{train} \vline \\
%\hline Steps  & 30 & 100 \\ %& 30 & 100 \\
%\hline Heap used (MB)& 10& 17 \\ %& 10 & 11 \\
%\hline Heap allocated (MB)& 63 & 63 \\ %& 63 & 63 \\
%\hline Parser (ms) & 750 & 750 \\ %& 800 & 800\\
%\hline Simulation (ms)& 200& 750 \\ %& 100 & 550 \\
%\hline
%\end{tabular}
%
%\end{center}
%\label{tab:profiler}
%%\begin{tablenotes}
%%{\scriptsize
%%\item Heap used (MB). Parser execution time (ms).  Simulator execution time (ms).
%%}
%%\end{tablenotes}
%\end{subtable}
%\end{table}

\smartref{tab:structures} shows the size of the store with respect to the number of steps  executed.
In the example, an increment of steps implies that more procedure calls and \texttt{exists} agents are executed. Consequently, the number of nodes in the symbol table increases with the number of steps executed and similarly the size of the global memory. In addition, when a new element is added to a stream, three registers are allocated: the functor, the head and the tail. Observe that the {\em photocopier} example has linear constraints (\texttt{Aux' = Aux -1}). Since the \texttt{photocopier} does not receive commands, the value stored in \texttt{T} is decremented until it is 0. In consequence, after some steps, $disc\_poly$ has 6 dimensions and the following constraint system: $$D\_0 = 5 \wedge D\_1 = 4 \wedge D\_2 = 3 \wedge D\_3 = 2 \wedge D\_4 = 1 \wedge D\_5 = 0 $$

We have used the \java\ monitoring console to collect information regarding the execution time and the memory allocated. The tests have been carried out in a virtualized Ubuntu machine, with 2GB of RAM memory and one processor. The \java\ virtual machine is OpenJDK Client VM 24.95-b01.
The bottom part of \smartref{tab:structures}~also shows the average heap usage and execution times obtained by executing several times the same example.
Observe that, the execution of the parser takes longer than executing 30 steps of the abstract machine. Obviously, when the number of steps is higher, the parser remains in the same magnitude, and the simulator consumes more time. The execution time has been measured without printing the symbol table and memory state, since printing slows down the  execution. For example, the simulator execution time of 30 steps printing the status of symbol table each step takes an average of 187 ms, which is more or less the execution time of the parser.
With respect to the memory, we have monitored the heap used and allocated. The memory used/allocated is not the real total memory used by the \java\ application. We should investigate how to obtain this information in future work.

\section{Related Work}
\label{sec:related}
In this section, we describe and compare other proposals in the literature for the implementation of concurrent, declarative or synchronous languages, some of which were mentioned in the Introduction.

 {\em Lustre}~\cite{Ha91} and {\em SIGNAL}~\cite{Pascalin95,GamatieG10} share with \tccp\ their declarative and synchronous character. Both languages are data-flow oriented, that is, programs operate over infinite sequence of values. They have been used for the modelling and analysis of industrial critical systems,
which proves that synchronous languages are not only useful in academia. For example, {\em Lustre} is the language underpinning a wide range of tools~\cite{Halbwachs05}, the most important being {\em SCADE Suite}~\cite{scade_web}, a toolset for modelling, simulating, verifying and generating certified code for critical systems. Similarly, {\em POLYCHRONY}~\cite{polychrony} is the development framework of {\em SIGNAL}. It provides mechanisms for the design, simulation, verification and code generation for distributed hardware platforms.

Unlike our proposal, the final aim of the tools developed for {\em Lustre} and {\em SIGNAL} is to generate code, optimized for specific platforms. Instead, the main concern of our architecture is its compositional character and,  in consequence, its ability to adapt to new constraint solvers, new language extensions, or new agent interpreters that execute the program differently. In fact, since the current implementation is based on an interpreted abstract machine running on \java, the performance completely depends on the underlying \java\ virtual machine.

In the context of concurrent logic programming, there are some older languages which share characteristics with \tccp. For example, {\em PARLOG}~\cite{Clark1986} is a logic concurrent  language descendent of {\em PROLOG}~\cite{Kowalski1988}, intended to describe distributed systems. {\em PARLOG} supports  fine-grain parallelism similar to that of \tccp. Process communication is also carried out using streams, but it lacks global memory and mechanisms for hiding and creating new local variables. In contrast, {\em PARLOG}  incorporates the notion of modes, associated with the process parameters, to synchronise them. In~\cite{Gregory1989}, the authors present an abstract machine for the implementation of {\em PARLOG} on uniprocessors. In this implementation, the {\em PARLOG} computation is represented by an AND/OR tree in which each node is a process that can be runnable or locked (waiting for some data), in a similar way to the \tccp\ agents. % Thus, the  execution state of a {\em PARLOG} program is defined by these runnable nodes.
Although,  both proposals (\tccp\ and {\em PARLOG}) agree that the agents (processes) have an associated sequence of instructions of the underlying abstract machine, the memory models are considerably different. This is principally because of the \tccp\ hiding operator which makes it possible to define an arbitrary number of local variables.
As regards implementation issues, both implementations provide simulators on the language, but use different target languages ({\em C} and {\em Java}).

Along the same lines, Kernel Language {\em KL1}~\cite{Ueda1990} is a committed-choice logic language based on the Guarded Horn Clauses {\em GHC}~\cite{Ueda1986} intended to be target language for the implementation of concurrent logic languages. The main goal of {\em KL1} is, therefore, the construction of  efficient, production-quality programs to exploit physical parallelism. The implementation of {\em KL1} was developed on the Parallel Inference Machine (PIM), which has a hierarchical memory-architecture where clusters have processing elements connected by a bus. In this implementation, the difficulties are also related to memory architecture and management.

With respect to \tccp\ tools, \cite{Sjoland01} presents an interpreter of \tccp\ implemented in Mozart-Oz~\cite{Haridi1998}. The semantics of \tccp\ is mapped into Mozart-Oz directly defining a translation from \tccp\ to Mozart-Oz. Nevertheless, this work is not publicly available and does not included the latest features of \tccp\ presented during the last years.

In~\cite{Lescaylle200963}, authors present a \tccp\ interpreter,  implemented in Maude. Similar to our proposal, they parse, interpret and simulate \tccp\ programs. Their tool implements six Maude modules, one for each \tccp\ entity (agents, constraints, programs, store, constraint system and operational semantics) which are used to directly translate from \tccp\ to Maude. %This proposal is its dependency on the Maude platform which is used as the interface to run and interpret the simulation results. In addition, all Maude modules implemented are strongly coupled, thus it is not possible to modify one of them independently.
%Alternatively, this application can be reused for implementing another \tccp\ tools, since it would only need to add the new modules that implement the new elements for the implementation without change anything in these six basic modules.

%Finally, it is important to note that there are other constraint solvers such as JaCoP~\cite{jacop_web}. This is a constraint solver which offers a large variety of constraints that include basic arithmetic operation constraints, logical and conditional constraints, regular constraints for the assignment to variables that are accepted by an automaton, etcetera. This constraint solver is versatile and easy to use in solutions such as this proposal. However, we have chosen the PPL as constraint solver because we plan to extend this tool for modelling hybrid systems which combines continuous and discrete variables and the PPL provides a great number of functionalities over continuous variables which allow us to represent this type of systems through of polyhedra and to operate over the polyhedra.
%
%

\section{Conclusions and Future Work}
\label{sec:conclusions}
%In this research, we have presented an abstract machine of Hy-tccp that is the basis of an implementation of Hy-tccp, an extension of tccp for hybrid systems \\

We have presented an abstract machine for \tccp, which defines the behaviour of \tccp\ agents over a memory architecture called {\em store}. The abstract machine is composed of different modules  which have been design to be as independent as possible. Most of the architecture components are unaware of the actual implementation of the memory or the particular implementation of the agent behaviour. We think that this approach facilitates and simplifies the development of tools for \tccp.
In addition, we have implemented a tool for the simulation of \tccp\ following this abstract machine architecture. %The tool implemented makes use of an interpreter for executing \tccp\ agents. Given a specification of a \tccp\ agent, the interpreter is able to execute agents following the language semantics.
%Making use of the abstract machine, we provide an implementation of a \tccp\ simulator.
The tool has been implemented in {\em Java}, and uses other external libraries and frameworks to implement different elements. For example, we use ANTLR to generate the parsers, and  \ppl\ to implement the constraint solver for linear constraints.

%We have also contributed a parser and an interpreter to Hy-tccp (and also to tccp). Both of them can be used to develop Hy-tccp and tccp tools of analysing, executing and verification.
%
%One of the important advantages of this implementation is that is based on formal verification methods for \tccp. We plan to develop semantics manipulation tools for the analysis and verification of \tccp\ programs using this implementation.

We have evaluated the simulator with the photocopier example, running different number of abstract machine steps. We have presented the state of the abstract machine store after executing the example, and shown some performance measures obtained with profiling tools.
We believe that the performance is acceptable, although we should improve the memory model to achieve more efficient implementations for constructing, for instance, a \tccp\ model checker.

Although, the current tool at \url{http://morse.uma.es/tools/tccp} may be only used to simulate  some available \tccp\ codes, we plan to extend its capability by allowing users to simulate their own programs. In fact, the tool only lacks a frontend that manages syntax errors.
In addition, due to the different implementation approaches followed by tool~\cite{Lescaylle200963} and ours, it is difficult to compare performance of both tools but we plan to do it in the near future.

As future work, we wish to extend the abstract machine to \hytccp~\cite{AdalidGT15}.
\hytccp\ is an extension of \tccp\ for hybrid systems, which adds a notion of continuous time and new agents to describe the continuous dynamics of hybrid systems.
\hytccp\ is independent from the kind of constraints over continuous variables. To implement a \hytccp\ simulator, we will assume that the hybrid systems are rectangular.
Because of the independence amongst the different entities which compose implementation, the extension of the current abstract machine will involve adding the new agents of the \hytccp\ language and probably new abstract machine instructions. In addition, the parser should be extended to recognise the new agents. Finally, we will reuse \ppl\ as the constraint solver for constraints over continuous variables.

%\begin{pleasenote}[MMar]si da tiempo mencionar lo de la interpretacion abstracta
%\end{pleasenote}

%Concerning to \hytccp\ and using a methodology similar to~\cite{GallardoP13}, we will adapt the \tccp\ model checking algorithms to \hytccp . In this way, we will be %able to analyse reachability, safety and other properties.

%Moreover, we ... \textbf{ABSTRACT INTERPRETATION \cite{Alpuente200558} and \cite{Cousot1977}}

%Although the current tool is being implemented in Java, the power of the abstract machine defined will allow us to explore some other possible implementations. For %instance, we also think to implement a \hytccp\ interpreter using Promela (the input language of model checker \cite{Hol04}) as target language. This new implementation will make it possible to carry out very efficient model checking algorithms on \hytccp\ programs.

%We have described an abstract machine for tccp that gives the general guidelines to develop an Hy-tccp implementation. In this way, we facilitate the development of tools for Hy-tccp and tccp, since the description of the agents behavior is defined in the abstract machine. \\

%In addition, we are developing a prototype tool based on the abstract machine. We are implementing an interpreter for Hy-tccp and tccp in Java. We have restricted the %constraints to linear constraints with integer coefficients. The main elements of the tool are: the parser, developed with ANTLR, the constraint solver, which is based on convex polyhedra and implemented with PPL, and finally, the implementation of the abstract store in java. \\

%\nocite{*}
\bibliographystyle{eptcsini}
\bibliography{biblio}
\end{document}